# Scattering from Star Polymers including Excluded Volume Effects

Xin Li,[†] Changwoo Do,[†] Yun Liu,[#,‡] Luis E. Sánchez-Diáz,[†] Kunlun Hong,[§] Gregory S. Smith,[†] and Wei-Ren Chen[*,†]

[†]Biology and Soft Matter Division, Oak Ridge National Laboratory, Oak Ridge, Tennessee 37831, United States

[#]The NIST Center for Neutron Research, National Institute of Standards and Technology, Gaithersburg, MD 20899-6100, United States

[‡]Department of Chemical Engineering, University of Delaware, Newark, DE 19716, United States

[§]The Center for Nanophase Materials Sciences, Oak Ridge, Tennessee 37831, United States

**ABSTRACT:** In this work we present a new model for the form factor of a star polymer consisting of self-avoiding branches. This new model incorporates excluded volume effects and is derived from the two point correlation function for a star polymer.. We compare this model to small angle neutron scattering (SANS) measurements from polystyrene (PS) stars immersed in a good solvent, tetrahydrofuran (THF). It is shown that this model provides a good description of the scattering signature originating from the excluded volume effect and it explicitly elucidates the connection between the global conformation of a star polymer and the local stiffness of its constituent branch.

Star polymers are synthetic macromolecules consisting of polymer branches emanating from the molecular center.[1-4] They are characterized by the structural features of both linear polymers and colloids. In addition to their scientific interest, they many industrial applications add to their importance.[1,5] Consequently they have been studied extensively.

Much of the contemporary knowledge of star polymers is the results of scattering experiments.[1-4] Small angle scattering provides a convenient tool for the structural investigation of star polymers over the spatial range of ~ 1 – 200 nm.[1-2,4] Numerous experimental results clearly demonstrate that the conformation of star polymers in solution is indeed different from that of linear polymers.[1-2,4] The additional intra-molecular spatial correlation, which does not exist in linear polymer chains, is reflected by an additional characteristic peak in the scattering from star polymers.

To quantitatively characterize the conformation of star polymers, theoretical models of scattering functions have been developed to understand experimental scattering studies. An analytical single particle scattering function $P(Q)$, often called the form factor, was first derived by Benoît.[6] In this model a star polymer is viewed as a collection of multiple ideal Gaussian chains with one end attached to the molecular core and the other free to move. This model shows quantitative agreement with the scattering results of star polymers dissolved in θ solvents[7] and immersed in melts[8]. However, since the excluded volume effect is not considered in this model, it fails to describe the scattering features of a star polymer with swollen chain statistics.[9-12] Later a phenomenological model was proposed by Dozier *et al.* to describe the sequential multi-step decay in the scattered intensity of star polymers dissolved in good solvents.[11] When compared with experiment satisfactory agreement is seen; however, detailed microscopic conformational information, such as the stiffness of the polymer branches and its connection to the overall polymer size cannot be explicitly obtained from this model. Although, theory has also been proposed to incorporate the influence of inter-branch interactions into the scattering function of a star.[13] there is a clear discrepancy when compared experiment.[1,9,11-12] To date a viable scattering function of star polymers with the explicit incorporation of excluded volume effects remains unavailable.

In this letter we derive such a scattering function theoretically. We demonstrate that our model indeed is able to describe the experimental scattering signature of star polymers originating from the excluded volume effect. There its availability facilitates the quantitative conformational characterization of this important class of soft matter.

The conformation of a star polymer is reflected by the two-point spatial correlation function, often called the form factor

$$P_{star}(Q) = \langle \exp[i\vec{Q} \cdot (\vec{r} - \vec{r}')] \rangle \qquad (1)$$

where the angle brackets represent the angular average, $\vec{r}$ and $\vec{r}'$ represent two arbitrary nuclear positions along the polymer chain. For a star polymer the positions $\vec{r}$ and $\vec{r}'$ in Eqn. (1) can either be located at the same branch or different ones. Therefore for a star polymer consisting of

$n_B$ branches, the corresponding form factor $P_{star}(Q)$ can be expressed as

$$P_{star}(Q) = P_{intra}(Q) + P_{inter}(Q)$$
$$= \frac{1}{n_B}\langle \rho(\vec{r})\rho(\vec{r}')\exp[i\vec{Q}\cdot(\vec{r}-\vec{r}')]\rangle_{intra}$$
$$+ \frac{n_B-1}{n_B}\langle \rho(\vec{r})\rho(\vec{r}')\exp[i\vec{Q}\cdot(\vec{r}-\vec{r}')]\rangle_{inter} \quad (2)$$

In Eqn. (2) $P_{intra}(Q)$ and $P_{inter}(Q)$ represent the scattering contributions of a single branch and the inter-branch spatial correlations respectively. The pre-factor $\frac{1}{n_B}$ denotes the possibility of finding positions $\vec{r}$ and $\vec{r}'$ on the same polymer branch, and $\frac{n_B-1}{n_B}$ for that on two different branches. $\rho(\vec{r}) = b_c \delta(\vec{r})$ gives the scattering length density (SLD) at position $\vec{r}$, where $b_c$ is the bound coherent scattering length of the nucleus at position $\vec{r}$.. Evaluation of Eqn. (2) requires the SLD distribution along the polymer branch. This distribution is governed by the end-to-end distance distribution function $p(r)$.[14-15] As indicated by its name, $p(r)$ gives the probability of finding the distance between two ends of a polymer chain with the value of $r$. For a star homopolymer with a constant $b_c$, Eqn. (2) can be expressed as

$$P_{star}(Q) = P_{intra}(Q) + P_{inter}(Q)$$
$$= \frac{1}{n_B}\langle \int d^3\vec{r}\, b_c^2 \exp[i\vec{Q}\cdot(\vec{r}-\vec{r}')] p^*_{intra}(\vec{r}-\vec{r}')\rangle_{intra}$$
$$+ \frac{n_B-1}{n_B}\langle \int d^3\vec{r}\, b_c^2 \exp[i\vec{Q}\cdot(\vec{r}-\vec{r}')] p^*_{inter}(\vec{r}-\vec{r}')\rangle_{inter} \quad (3)$$

Where

$$p(r) = 4\pi r^2 p^*(\vec{r}) \quad (4)$$

In Eqn. (3), the term of $b_c^2$ resulting from the two-point spatial correlation is set to be 1. For an ideal Gaussian chain, it is expected that $p_{intra}(r) = p_{inter}(r)$ because the excluded volume effect is neglected both between different chains and the monomers within the same chain. In this approximation, $p(r)$ is given by a Gaussian function and $P_{intra}(Q)$ and Eqn. (3) reduce to the Debye function[16] and Benoît function[6] respectively.

Let's consider how to include the excluded volume effects. Figure 1 shows a schematic representation of a star polymer immersed in good solvent. With an increase in the number of branches, $n_B$, the average accessible space decreases progressively due to the enhanced excluded volume effect. It has been demonstrated computationally that $p(r)$ of each branch is shifted towards the molecular periphery.[17] Therefore with the addition of a branch, each branch is confined within a cone characterized by a decreasing solid angle $\Omega$. As a result each branch takes a more stretched conformation. Rigorously, $p_{inter}(r)$ presented in Eqn. (3) can only be obtained by providing the expression of the center-to-end distance distribution of the two partial chains with $n$ and $n'$ monomers, and the angle $\theta$ between them. To date the analytical expression of this center-to-end distance distribution as a function of $n_B$ remains unknown. Nevertheless if $n_B$ is sufficiently small, the inter-branch excluded volume effect may not affect the conformation of each branch drastically. Accordingly one can assume that the conformation of a joint chain made of two connecting partial branches covalently bound at the center (marked by red color in Figure 1) is governed by the same end-to-end length distribution function of a self-avoiding polymer chain. In this model any two jointed partial branches is approximated as a single chain. With this assumption, both $p_{intra}(r)$ and $p_{inter}(r)$ can take the same end-to-end length distribution function $p(r)$ of a free single chain. Consequently Eqn. (3) can be further written as

$$P_{star}(Q) = P_{intra}(Q) + P_{inter}(Q)$$
$$= \frac{1}{n_B}\frac{1}{N^2}\int_0^N dn \int_0^N dn' \int_0^\infty dr\, \frac{\sin(Qr)}{Qr}p(|n-n'|,r)$$
$$+ \frac{n_B-1}{n_B}\frac{1}{N^2}\int_0^N dn \int_0^N dn' \int_0^\infty dr\, \frac{\sin(Qr)}{Qr}p(n+n',r) \quad (5)$$

where $N=L/b$, and $L$ and $b$ are the contour length and the length of a Kuhn segment, respectively. $N$ characterizes the stiffness of a single branch. The smaller $N$ becomes, the stiffer the branch is. Therefore, if $p(r)$ is specified, $P_{star}(Q)$ can be obtained numerically as a function of $n_B$, $L$ and $b$. In our model the mathematical expression of $p(r)$ proposed by des Cloizeaux[18-20] is chosen because of its numerical accuracy as demonstrated by computer simulation.[21-22] Moreover, a parameterized expression of $P_{intra}(Q)$[23-24] is used to further enhance the computational efficiency.

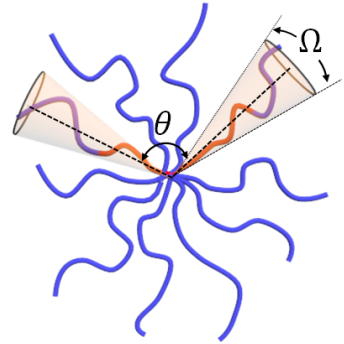

Figure 1. Schematic representation of a star polymer. Each branch is confined within a space subtended by a solid angle $\Omega$. Two partial chains with $n$ and $n'$ monomers are marked with red color. $\theta$ represents the angle between them.

In Figure 2 we plot the $P(Q)$ versus Q calculated from the Benoît function[6] (Figure 2(a)) and our model given in Eqn. (5) (Figure 2(b)). The conformational parameters, $n_B$, $L$, and $b$, used in the both calculations are 15, 100 Å and 20 Å respectively. For a Gaussian star its $P(Q)$ (Figure 2(a)) resembles that of linear polymer chains. Beyond the Guinier region ($Q \ll 0.01$ Å$^{-1}$), the intensity of $P(Q)$ falls off monotonically. Existing experimental results have already demonstrated that the Benoît function is unable to describe the scattering signature of star polymers immersed in good solvents. On the contrary, the $P(Q)$ calculated from Eqn. (5) (Figure 2(b)) is seen to exhibit a characteristic deflection occurring around $Q = 0.1$ Å$^{-1}$. Qualitatively this observation is consistent with the $P(Q)$ for a star polymer, obtained from both experiments[11-12,26-31] and simulations[17,25], including excluded volume effects. There have been discussions about the connection between the excluded volume effects and the scattering signature of star polymers under the good solvent conditions. An insightful perspective is gained from our model calculation: In comparison to its counterpart presented in Figure 2(a), the $P_{inter}(Q)$ presented in Figure 2(b) appears to decay faster beyond the Guinier region because of the excluded volume effect incorporated via the des Cloizeaux approximation. Meanwhile, as demonstrated in Figure 2(a), the $P_{intra}(Q)$ for a random walk polymer, which follows the Debye function, exhibits a $Q^{-2}$ dependence in the high-$Q$ Porod region. On the other hand it has been known[15,32] that the excluded volume interactions between the constituent monomers causes the swelling of a polymer branch. As a result the corresponding $P_{intra}(Q)$ (dashed line of Figure 2(b)) deviates from the Debye function and follows a different power law in the Porod region, whose specific form depends on experimental conditions.[25] Therefore our calculation explicitly demonstrates that the excluded volume effect causes the deviations of $P_{inter}(Q)$ and $P_{intra}(Q)$ from the ideal condition, and consequently produces the observed two-step decay of $P(Q)$.

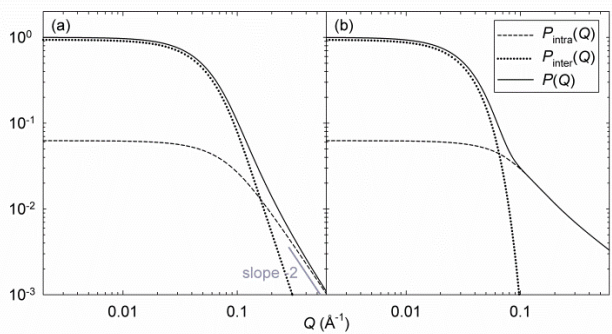

Figure 2. $P(Q)$ of a star polymer calculated from (a) Benoît function and (b) Eqn. (5). The conformational parameters, $n_B$, $L$, and $b$, used in the calculations are 15, 100 Å and 20 Å respectively. $P_{intra}(Q)$ and $P_{inter}(Q)$ represent the scattering contributions from the single chain and inter-chain spatial correlations.

A natural question is then, upon changing the conformational variables, if our model will be able to reproduce the same experimentally measured characteristic variations of $P(Q)$. Several investigations has been devoted to understanding how $P(Q)$ of a star polymer immersed in a good solvent evolves as a function of $n_B$.[12,17,25-26,29-31] Upon increasing $n_B$, both experiment and computer simulations found that the two-step falling off of the scattering intensity becomes progressively pronounced. As demonstrated by Figure 3(a), this intensity variation can be well described qualitatively by our model. Based on Eqn. (5), one can attribute the origin of the evolving $P(Q)$ to the increasing contribution of $P_{inter}(Q)$, which reflects the increasing compactness of a star polymer. Again this observation is consistent with the intuitive expectation as a result of enhancing intra-star steric hindrance. The impact of solvent quality on $P(Q)$ has also been explored.[17,31] Upon varying the solvent quality from good to θ condition, the decay of $P(Q)$ is seen to be pushed towards the higher-$Q$ region and its two-step variation in the Porod region is seen to become less pronounced. It has been known that a self-avoiding polymer chain takes a more stretched conformation in better solvents.[14-15,32-33] By increasing the ratio of $L/b$, the influence of deteriorating solvent quality on $P(Q)$ can be directly explored by our model. Again the same qualitative trend is seen in Figure 3(b).

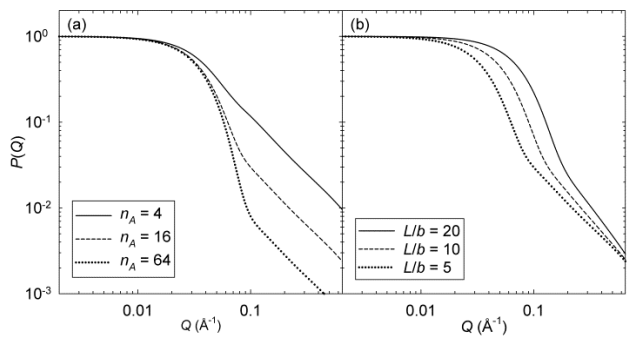

Figure 3. (a) $P(Q)$ of a star polymer calculated respectively from Eqn. (5) as a function of branch number $n_B$. $L$ and $b$ used in the calculations are 100 Å and 20 Å respectively. (b) $P(Q)$ of a star polymer calculated respectively from Eqn. (5) as a function of $L/b$. $n_B$ used in the calculations is 15.

Having seen that Eqn. (5) produces scattering functions which are qualitatively consistent with documented results under a variety of conditions, one can readily examine the viability of using this model to extract the conformational properties of star polymers from experimental data. In Figure 4 we present the modified Kratky plot[34] of small angle neutron scattering (SANS) absolute intensity (symbols) of 15-branches polystyrene (PS, $C_8H_8$) star polymers immersed in fully deuterated tetrahydrofuran (d8-THF, $C_4D_8O$), along with the model fit (solid curves). The PS star polymers and d-THF are were obtained from Polymer Source Inc. and Cambridge Isotope Laboratories Inc. respectively.[35] The molecular weight of each PS branch is 1.1 kDa and the total molecular weight of a PS star is 17 kDa. The concentration of PS star polymers dissolved in d-THF was kept at 0.5 wt% to avoid the possible coherent scattering contribution from inter-molecular spatial correlations. SANS measurements were performed at the NG-7 SANS instruments at the NIST Center for Neutron Research (NCNR) and at the EQ-SANS instrument at Spallation Neutron Source (SNS) ORNL. For the SANS measurement at NCNR, the wavelength of the incident neutrons was 6.0 Å, with wavelength spreads of 15%. The scattering wave vector $Q$ ranges from 0.004 to 0.4 Å$^{-1}$. For the SANS measurement at SNS, three configurations of sample-to-detector distance of 4 m, 2.5m, and 1.3m with two neutron wavelength bands were used to cover the $Q$ range of 0.005 to 0.4 Å$^{-1}$. The measured scattering intensity was corrected for detector sensitivity and the background from the empty cell, and placed on an absolute scale using a calibrated standard. Both SANS measurements were carried out using Hellma quartz cells of 1 mm path length at 25.0 ± 0.1 °C. At this temperature THF is a good solvent for PS. , Excellent agreement is seen between our model and SANS experiment. The $L$ and $b$ extracted from the model fitting are 75.8 ± 3.1 Å and 8.83 ± 0.43 Å respectively.The extracted conformational parameters are physically reasonable given the molecular weight of each branch and the solvent condition, . From the results presented in the Kratky plot, the coherent scattering contributions from the inter- and intra-branch spatial correlations, as expected, are mainly manifested in the low-$Q$ Guinier region and high-$Q$ Porod region respectively.

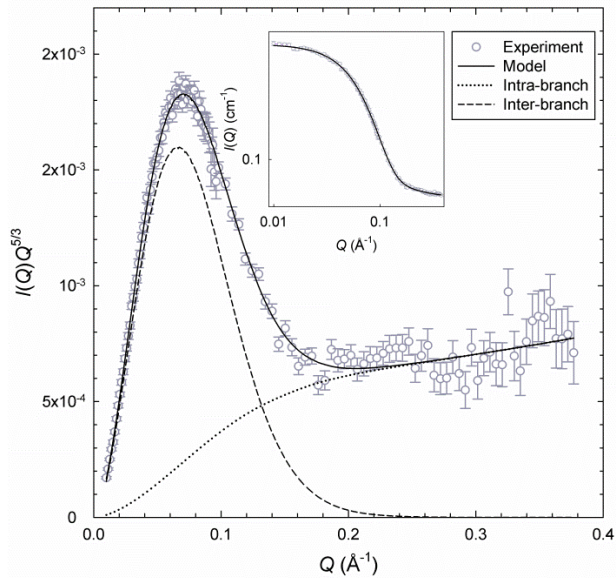

Figure 4. The Kratky plot of the SANS coherent scattering intensity $I(Q)$ obtained from the 15-branches PS star polymers dissolved in fully deuterated THF solvent with 0.5 wt% concentration. The solid, dotted and dashed lines are the model fitting, and the scattering contribution from intra- and inter-branch spatial correlations. Inset give the SANS $I(Q)$ along with the model fitting.

It is worth comparing our model scattering function with the one proposed by Dozier and coworkers based on a phenomenological approach.[11] First of all, both models exhibit quantitative agreements with experimental results. However, instead of directly deriving the $P(Q)$ from the two-point spatial correlation function (Eqn. (1)), in the model by Dozier the coherent scattering contribution from the length scales corresponding to the Guinier and Porod regions is first derived separately, and later joined together to render the $P(Q)$ via a linear summation. Because this model is developed compositely, the fitting parameters representing the conformational properties at different length scales are essentially independent of each other. For example, a direct mathematic connection between the radius of gyration $R_G$ and the blob size $\xi$, the main fitting parameters for the Guinier and Porod regions, does not exist. Moreover, in this model the scattering contribution from the Porod region is derived based on a scaling picture. How to relate the variation of $\xi$ obtained from model fitting to any conformational changes at microscopic level is not intuitively obvious. On the other hand, since our model example is directly derived from Eqn. (1), the conformational parameters at different length scales can be related to each other. For example, in our model $R_G$ can be defined based on the mean-squared distance (MSD)[2]

$$\langle R_G^2 \rangle = \frac{1}{2\left(\frac{L}{b}n_B\right)^2} \sum_{i,j} \langle (r_i - r_j)^2 \rangle \qquad (6)$$

Since $\sum_{i,j} \langle (r_i - r_j)^2 \rangle$ can be evaluated numerically given the des Cloizeaux approximation, $R_G$ can be calculated accordingly. Nevertheless, it is critical to point out the validity and limitation of our proposed model. Because the approximations made for simplifying Eqn. (4) to Eqn. (5), there is no doubt that this model only valid for describing the conformational characteristics of a star polymer with low number of polymer branches under good solvent conditions. Any detailed description of star polymers with high number of branches certainly requires a modification and improvement of current model. At present, theoretical studies, complemented by computer simulations and SANS experiment, are underway in our laboratory.

To summarize, we report a theoretical model for the scattering function of star polymers under good solvent conditions. This model is directly derived from the two-point spatial correlation function with the incorporation of the excluded volume effect via the des Cloizeaux approximation. Excellent agreement is found between our model and SANS experiments using physically reasonable conformational parameters. In comparison to existing models which are derived phenomenologically, one advantage of our model is that it describes the conformation of a star polymer mechanistically based on three primary parameters including the number of branches $n_B$, the contour length of each branch $L$ and the Kuhn length $b$ which measures the stiffness of a constituent branch. This model is shown to be valid for low concentrations of star polymers in good solvent conditions.

## AUTHOR INFORMATION

**Corresponding Author**

*E-mail: chenw@ornl.gov

## ACKNOWLEDGMENT

This work was supported by the U.S. Department of Energy, Office of Basic Energy Sciences, Materials Sciences and Engineering Division. The EQSANS experiment at the SNS at Oak Ridge National Laboratory was sponsored by the Scientific User Facilities Division, Office of Basic Energy Sciences, U.S. Department of Energy. We also acknowledge the support of the National Institute of Standards and Technology, U. S. Department of Commerce, in providing the neutron research facilities used in this work.

SYNOPSIS TOC (Word Style "SN_Synopsis_TOC"). If you are submitting your paper to a journal that requires a synopsis graphic and/or synopsis paragraph, see the Instructions for Authors on the journal's homepage for a description of what needs to be provided and for the size requirements of the artwork.

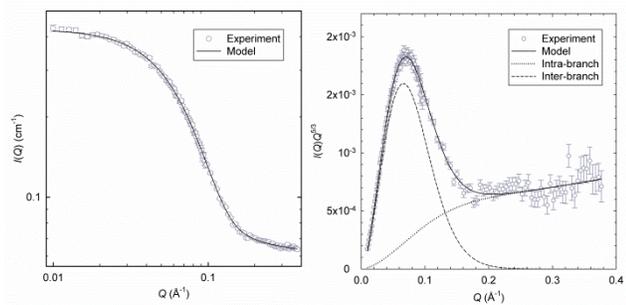